# The Ferroelectric Superconducting Field Effect Transistor

Alessandro Paghi[1*], Laura Borgongino[1], Elia Strambini[1], Giorgio De Simoni[1], Lucia Sorba[1], and Francesco Giazotto[1]

[1]Istituto Nanoscienze-CNR and Scuola Normale Superiore, Piazza San Silvestro 12, 56127 Pisa, Italy.

[*]Corresponding authors: alessandro.paghi@nano.cnr.it

## Abstract

The ferroelectric field-effect transistor (Fe-FET) is a three-terminal semiconducting device first introduced in the 1950s. Despite its potential, a significant boost in Fe-FET research occurred about ten years ago with the discovery of ferroelectricity in hafnium oxide. This material has been incorporated into electronic processes since the mid-2000s. Here, we observed ferroelectricity in a superconducting Josephson FET (Fe-JoFET) operating at cryogenic temperatures below 1 Kelvin. The Fe-JoFET was fabricated on the InAsOI platform, which features an InAs epilayer hosted by an electrical insulating substrate, using $HfO_2$ as the gate insulator, making it a promising candidate due to its ferroelectric properties. The Fe-JoFET exhibits significant hysteresis in the switching current and normal-state resistance transfer characteristics, which depend on the range of gate voltages. This phenomenon opens a new research area exploring the interaction between ferroelectricity and superconductivity in hybrid superconducting-semiconducting systems, with potential applications in cryogenic data storage and computation. Supporting this, the Fe-JoFET was operated as a cryogenic superconducting single memory cell, exhibiting both dissipative and non-dissipative states. Its non-volatility was tested over a 24-hour measurement period. We also demonstrated that the Fe-JoFET can retain information at temperatures above the superconductor critical temperature, resulting in a temperature-fault-tolerant memory cell resistant to temperature oscillations or, in the worst case, cryostat faults.

**Introduction**

The ferroelectric field-effect transistor (Fe-FET) is a well-known three-terminal semiconductor device first introduced in the 1950s [1]. It incorporates a ferroelectric insulator within the gate dielectric stack, causing characteristic hysteresis in its electrical behavior. The Fe-FET can serve as a non-volatile memory cell, where data is encoded in the direction of ferroelectric polarization. The latter influences the formation of the conductive channel in the semiconductor layer, leading to opposite shifts in the threshold voltage of the Fe-FET [2].

Despite its promise, the widespread adoption of Fe-FETs has been hindered for over five decades due to integration challenges, particularly those involving perovskite-based ferroelectric oxides, which, although effective in ferroelectric random-access memory (FeRAM), pose challenges to incorporating them into standard semiconductor manufacturing processes [2]. The discovery of ferroelectricity in hafnium oxide in 2011 marked a turning point, triggering a surge in Fe-FET research [3]. Hafnia, already a cornerstone of high-$k$ metal gate (HKMG) technology used in advanced logic transistors since the 45nm technological process in 2007, offers compatibility with modern complementary metal-oxide-semiconductor (CMOS) processes and scalability [4]. This makes it a promising material for realizing Fe-FETs in high-volume semiconductor manufacturing. A significant milestone was reached in 2017, when a Fe-FET was successfully integrated into a 22nm fully-depleted silicon-on-insulator (FDSOI) CMOS process as a non-volatile single memory cell [5].

While Fe-FETs have shown great promise for non-volatile memory applications at room temperature, their potential extends beyond conventional computing systems. With growing interest in cryogenic electronics [6–11], driven by breakthroughs in superconducting and quantum technologies [12–20], there is a rising demand for memory devices that can operate reliably at ultra-low temperatures [21]. The reduced number of cryogenic memory technologies capable of operating at 4 K represents a critical challenge that limits the scalability and practicality of cryogenic electronic systems [21][22]. The scenario worsens if a sub-Kelvin operating temperature is required. Specifically, in a quantum computer, the control processor is connected to the quantum processing units, where qubits are located, using high-density superconducting interconnects to minimize the thermal leakages. Then, wires are used to connect the control processor to the room temperature electronics and memory data storage, with a restricted data bandwidth related to the significant wiring length. For this reason, cryogenic memories are an enabling technology for the achievement of a scalable quantum computing architecture [23][24].

Here, we observed the ferroelectric behavior of a hybrid superconducting Josephson FET (JoFET) operating at a cryogenic sub-K temperature. The resulting device, called Ferroelectric JoFET (Fe-JoFET), was fabricated on the InAsOI hybrid platform, where an InAs epilayer was grown onto a cryogenic electrical insulating substrate [25]. The device employs $HfO_2$ as the gate insulator, which introduces ferroelectricity. The Fe-JoFET exhibits pronounced hysteresis in both the switching current and normal-state resistance versus gate voltage transfer characteristics. Interestingly, the Fe-JoFET can be refreshed to its pristine state on demand by applying a tailored ambipolar gating protocol. We demonstrated that the Fe-JoFET can be used as a cryogenic superconducting non-volatile single memory cell, maintaining excellent state retention and readout stability over 24 hours of continuous measurement.

**Results and Discussion**

Figures 1a,b show the 3D and the cross-section structures of the JoFET fabricated on the InAs on Insulator (InAsOI) platform [25].

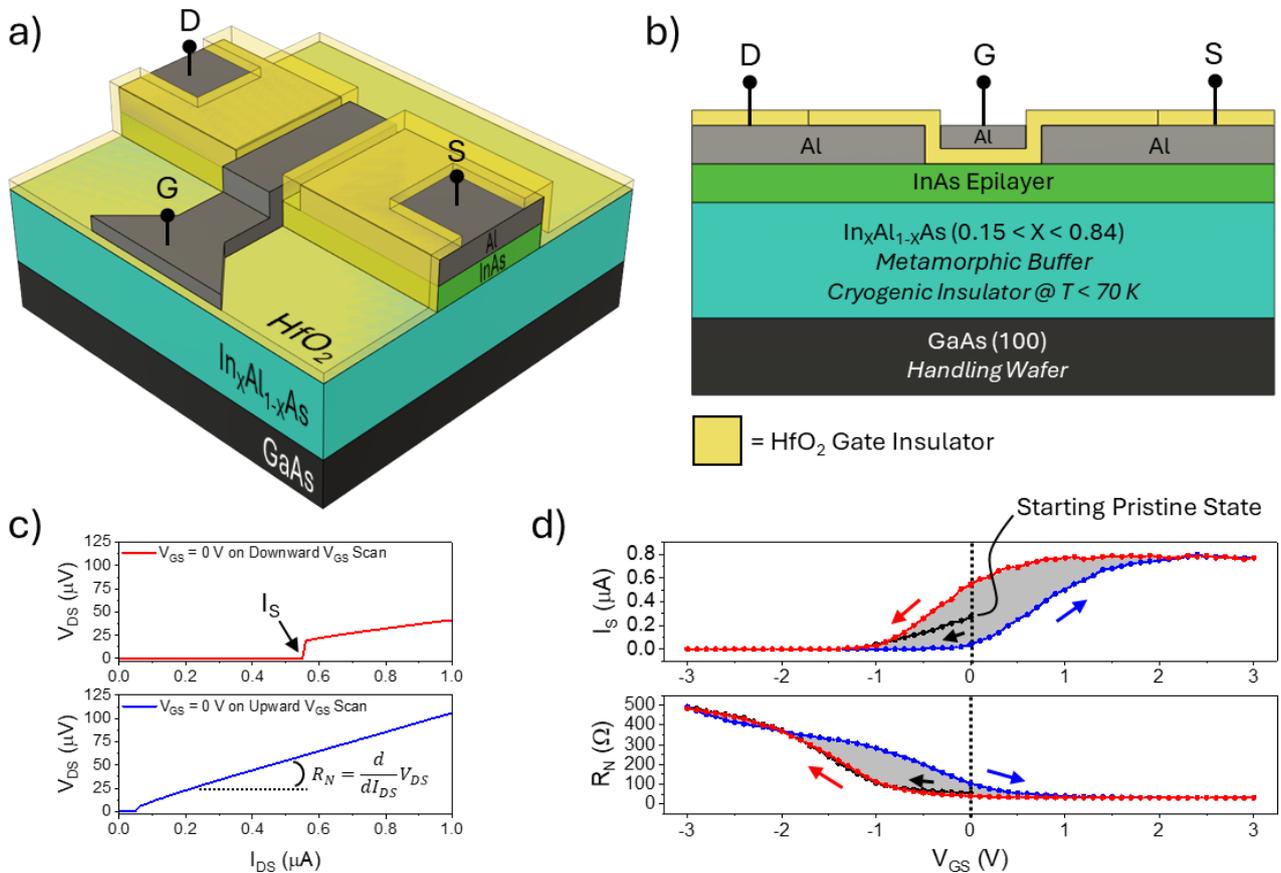

**Figure 1: Ferroelectric hysteresis in the InAsOI-based Josephson Field Effect Transistor.** a) 3D and b) cross-sectional views of an InAsOI-based JoFET. c) Voltage versus current curve of the InAsOI-based JoFET at $V_{GS}$ = 0 V during both downward (top) and upward (bottom) $V_{GS}$ scans. d) Switching current (top) and normal-state resistance (bottom) versus gate voltage transfer

characteristic recorded with a continuous $V_{GS}$ routine starting from a non-polarized pristine state of the JoFET. Panels (c, d) refer to an InAsOI-based JoFET with $W_{JJ}$ = 5 μm, $L_{JJ}$ = 600 nm, and $L_G$ = 600 nm measured at a bath temperature of 50 mK.

The InAsOI stack was realized via Molecular Beam Epitaxy and consists of a 500 μm-thick semi-insulating GaAs (100) substrate, a ~1.5 μm-thick step-graded $In_XAl_{1-X}As$ metamorphic buffer with X increasing from 0.15 to 0.84, and a 100 nm-thick intrinsically *n*-type semiconducting InAs epilayer [25]. The InAs exhibits a sheet electron density ($n_{2D}$) of $1.4×10^{12}$ cm$^{-2}$, a sheet resistance ($\rho_{2D}$) of 680 Ω, and a mobility ($\mu_n$) of $6.7×10^3$ cm$^2$/Vs at 3 K. The InAlAs metamorphic buffer layer exhibits electrical insulating behavior at temperatures below 70 K [25]. An Al layer (100 nm thick) was used as the superconductor for the source and drain leads, while the gate stack comprises a $HfO_2$ insulating layer (30 nm thick) and an Al layer (60 nm thick). The $HfO_2$ layer was grown via Atomic Layer Deposition (ALD) and characterized in previous runs using metal-insulator-metal capacitors, yielding a relative permittivity of 16.5 and a dielectric strength of 370 MV/m at 3 K [26]. We fabricated an InAsOI-based JoFET with a width ($W_{JJ}$) of 5 μm, interelectrode separation ($L_{JJ}$) of 600 nm, and gate length ($L_G$) of 600 nm. The JoFET was measured at 50 mK in a dilution cryostat equipped with a DC measurement. Figure 1c shows V-I curves of the JoFET at $V_{GS}$ = 0 V, reached during downward and upward *continuous* $V_{GS}$ scans, for which a triangular waveform $V_{GS}$ was used with amplitude continuously decreasing then increasing with time in the range [-3,3] V. The switching current ($I_S$), i.e., the maximum non-dissipative current supported by the JoFET, and the normal-state resistance ($R_N$), evaluated from the slope of the V-I curve above $I_S$, were highlighted on the graph for clarity. From the comparison between the two V-I characteristics, it is evident that the behavior of the JoFET at $V_{GS}$ = 0 V depends on the history of $V_{GS}$, both for $I_S$ and $R_N$. During the downward scan, the JoFET features an increased $I_S$ and reduced $R_N$ compared to those observed during the upward scan. Figure 1d summarizes the $I_S$ vs. $V_{GS}$ and the $R_N$ vs. $V_{GS}$ transfer characteristics retrieved from the V-I curves recorded during the $V_{GS}$ scans, with the JoFET starting from a non-polarized pristine state (block dots). The JoFET exhibits sigmoidal behavior in the collected trends: as $V_{GS}$ increases during the upward scan, $I_S$ increases while $R_N$ decreases. Similarly, decreasing $V_{GS}$ during the downward scan results in a reduction of $I_S$, accompanied by an increase in $R_N$. This behavior is consistent with the modulation of charge in the proximitized InAs epilayer, exhibiting an *n*-type polarity. Notably, the JoFET can operate in both enhanced (at $V_{GS}$ > 0 V) and depletion (at $V_{GS}$ < 0 V) modes. The observed counterclockwise and clockwise hysteresis in the $I_S(V_{GS})$ and $R_N(V_{GS})$ characteristics (Figure S1a), respectively, aligns with the typical behavior of n-type ferroelectricity exhibited by the $HfO_2$ gate insulator layer [27][28]. In contrast, alternative hysteretic mechanisms, such as electron trapping at the semiconductor/gate insulator interface, would produce an opposite trend, as commonly observed in similar non-ferroelectric devices (Figure S1b) [29].

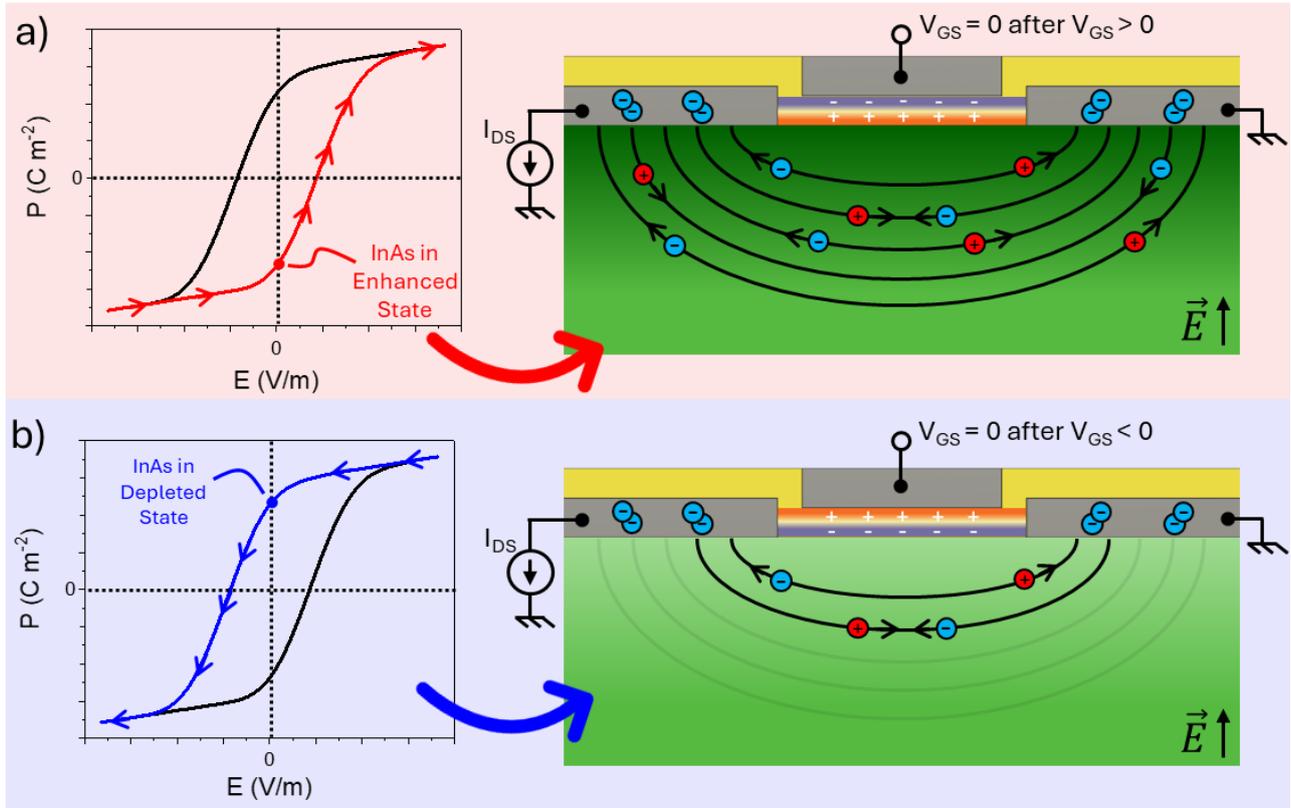

**Figure 2: Ferroelectricity at work in the InAsOI-based JoFET.** Microscopic sketch to explain how the gate insulator ferroelectricity acts on the JoFET electrical performance. a) Increasing the gate electric field from negative values results in a negative remanent polarization of the insulator at zero electric field, with the consequent enhancement of electrons in the InAs semiconducting layer, which can contribute to the supercurrent. b) A positive remanent polarization of the insulator, which arises when the gate electric field is decreased from positive values to zero, results in the depletion of electrons in the InAs layer and a reduction in the supercurrent.

Specifically, thanks to the charge trapping activated by positive $V_{GS}$ values, the depletion of electrons from the InAs epilayer into interfacial traps results in a reduced switching current and increased normal-state resistance during the downward $V_{GS}$ scan. The opposite happens during the upward $V_{GS}$ scan. Figure 2 depicts the microscopic explanation of how the gate insulator ferroelectricity acts on the JoFET electrical performance. The ferroelectric hysteresis of the polarization ($P$) vs. electrical field ($E = -\frac{V_{GS}}{d}$, where $d$ is the gate insulator thickness) curve leads to two different polarization states for the gate insulator. Increasing (with sign) the gate electric field from a high negative value results in a negative remanent polarization of the insulator at zero electric field, with the consequent enhancement of electrons in the InAs semiconducting layer (Figure 2a). The higher the number of majority carriers, the higher the supported supercurrent and, at the same time, the lower the normal-state resistance. On the other hand, a positive remanent polarization of the insulator arises at zero applied electric field when the gate electric field decreases from high positive values (Figure 2b). The InAs layer was subsequently subjected to a negative electric field, resulting in electron depletion in the JoFET channel and a reduction in supercurrent, accompanied by an increase in normal-state resistance. The ferroelectric behavior of the ALD-grown $HfO_2$ is well reported in the literature when subjected to a during-growth doping [30][31]. Interestingly, we observed that $HfO_2$ films grown by ALD in our chamber exhibited ferroelectric behavior even without intentional doping, but only when paired with a pristine Al gate. In contrast, devices incorporating Au or Ti/Al gate electrodes did not exhibit ferroelectricity [26][29]. Although unintentional doping during the ALD process cannot be completely ruled out, this suggests a chemically active role of the Al gate in stabilizing the ferroelectric phase, possibly through interfacial interactions or diffusion. Despite the ferroelectric origin, this device marks the first observation of ferroelectricity in a JoFET, and it is called the Ferroelectric JoFET (Fe-JoFET).

Figure 3a reports $I_S$ vs. $V_{GS}$ and $R_N$ vs. $V_{GS}$ characteristics of the Fe-JoFET gated with a *continuous $V_{GS}$* routine (Figure 3a inset) in the range [-1,1] V, [-2,2] V, and eventually [-3,3] V. The increase in the $V_{GS}$ range clearly enhanced the hysteretic behavior of the Fe-JoFET. For all the curves, is shown the second cycle recorded in a row with the Fe-JoFET starting from a non-polarized pristine state. This condition was reached by using an *ambipolar* $V_{GS}$ scan to refresh the Fe-JoFET, for which an alternate sign $V_{GS}$ was used with magnitude first increasing then decreasing with time in the range [-3,3] V (Figure 3b inset) [32]. Figure 3b reports the transfer characteristics of the Fe-JoFET gated with the *ambipolar $V_{GS}$* routine. There is no hysteresis between the $I_S$ and $R_N$ values obtained by increasing the $V_{GS}$ in the module and those achieved by decreasing the $V_{GS}$ in the module. This is because values achieved with the *continuous $V_{GS}$* routine represent the $I_S$ and $R_N$ values obtained at the extreme points of hysteresis curves for each specific $V_{GS}$ range starting from a pristine non-polarized state of the JoFET. Since the extreme points of a hysteresis curve do not exhibit hysteresis, there is also no hysteresis in the transfer characteristics obtained with the *ambipolar* $V_{GS}$ routine. We note with interest that, regardless of the ambipolar range (Figure S2a) and of the step size (Figure S2b), no differences are observed in the transfer characteristics achieved with the *ambipolar* $V_{GS}$ routine. The proposed method represents an excellent way to reset the Fe-JoFET to its pristine state, erasing the effect of the transistor history induced by any user activity. We want to emphasize that, to achieve the pristine state of the Fe-JoFET, a small step size should be used to suppress any hysteresis induced by reaching the 0 V bias. The *ambipolar* $V_{GS}$ routine was demonstrated as an effective method to recover the pristine state of JoFETs when the transfer characteristic hysteresis is dominated by the charge-trapping mechanism (Figure S3).

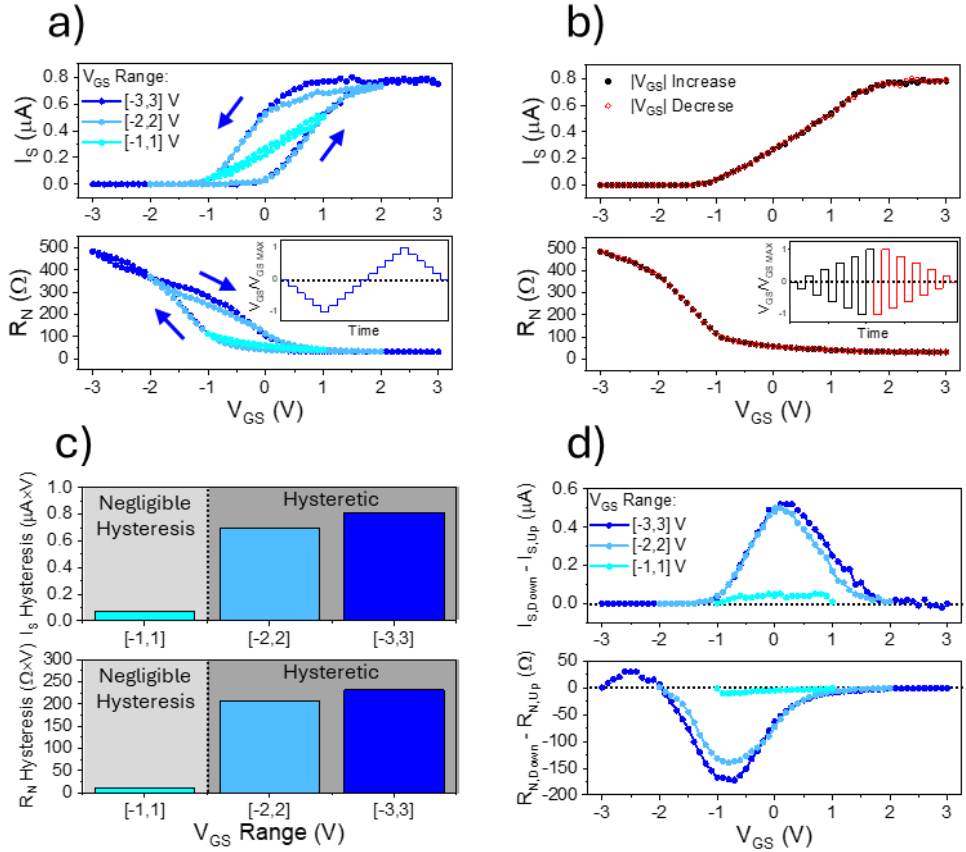

**Figure 3: Gate control of hysteresis in the InAsOI-based Fe-JoFET.** a) Switching current (top) and normal-state resistance (bottom) vs. gate voltage transfer characteristics for different gate voltage ranges recorded with the *continuous* $V_{GS}$ routine. The inset shows the *continuous* $V_{GS}$ routine. b) Switching current (top) and normal-state resistance (bottom) vs. gate voltage transfer characteristic recorded with the *ambipolar* $V_{GS}$ routine. The inset shows the *ambipolar* $V_{GS}$ routine. c) Switching current (top) and normal-state resistance (bottom) hysteresis area vs. gate voltage ranges retrieved from (a). d) Difference between downward and upward scans in switching current (top) and normal-state resistance (bottom) vs. gate voltage ranges retrieved from (a). a,b) For all the curves, the cycle shown is the second cycle recorded in a row with the Fe-JoFET starting in a non-polarized state. The measurement bath temperature is 50 mK.

Figures 3c,d report a detailed analysis of the Fe-JoFET hysteresis as a function of the gate voltage range. Figure 3c shows the $I_S$ and $R_N$ hysteresis area, evaluated as the integral area between downward and upward curves (Figure 1d). When the $V_{GS}$ range is increased from [-1,1] to [-3,3] V, both the hysteresis areas significantly rise from 0.07 to 0.8 μA×V for $I_S$ and from 10 to 231 Ω×V $R_N$, respectively. In the $V_{GS}$ range [-1, 1] V, the Fe-JoFET exhibits negligible hysteresis, indicating stable operation with minimal memory effect. However, expanding the $V_{GS}$ range to [-2, 2] V and higher induces a pronounced hysteresis, signifying the activation of ferroelectric polarization. This behavior implies that, once the Fe-JoFET is polarized using a $V_{GS}$ range of at least [-2, 2] V, it can subsequently operate within the reduced range of [-1, 1] V without disturbing the encoded polarization state. Notably, this enables a 50% reduction in the voltage required for non-polarizing (read) operations compared to the polarizing (write) phase, thereby enhancing energy efficiency. This feature allows for the reduction of the voltage required to write and read the Fe-JoFET. Figure 3d reports the difference between the downward and upward scan values, which enables the identification of the location of the maximum hysteresis point. The difference in the $I_S$ scans exhibits a peaked form, with the maximum value located at 0.1 V regardless of the gate voltage range used. Similarly, the difference in the $R_N$ scans also exhibits a peaked form, with the maximum value achieved at -0.8 V.

The hysteresis related to the ferroelectric behavior of the gate insulator allows us to define two distinct operating states of the Fe-JoFET at the same $V_{GS}$ value, i.e., *state 0* and *state 1*, thereby envisioning the use of the proposed device as a cryogenic superconducting single memory cell. In the following, we will focus on using the $V_{GS}$ range [-3, 3] V with the *continuous $V_{GS}$* gating routine, as it enhances the Fe-JoFET hysteresis (Figure 4a). $V_{GS}$ = 3 V ($V_{WRITE,0}$) and $V_{GS}$ = -3 V ($V_{WRITE,1}$) were used to set the Fe-JoFET in *states 0* and *1*, respectively. On the other hand, since the $I_S$ hysteresis peak was located near 0 V (Figure 3d), $V_{GS}$ = 0 V ($V_{READ}$) was used to read the encoded state of the Fe-JoFET. In *state 0*, the Fe-JoFET features the higher value of $I_S$ ($I_{S,0}$ = 536 ± 8 nA) and lower value of $R_N$ ($R_{N,0}$ = 41.0 ± 0.3 Ω), while in *state 1*, the Fe-JoFET exhibits the lower value of $I_S$ ($I_{S,1}$ = 33 ± 4 nA) and higher value of $R_N$ ($R_{N,1}$ = 110.1 ± 1.2 Ω). Previous data are retrieved over 10 writing and reading cycles, where a maximum coefficient of variation (CV = standard deviation / average value) of 12% is calculated, which serves as a clear benchmark for the reproducibility of the encoding operation.

Figure 4b exhibits the operating usage of the programmable Fe-JoFET. By using a reading drain-to-source current ($I_{READ}$) such that $I_{S,1} < I_{READ} < I_{S,0}$, in the *state 0*, the Fe-JoFET is in the non-dissipative state with a 0 Ω path between source and drain terminals. Conversely, in *state 1*, the Fe-JoFET is dissipative, exhibiting the resistance $R_{N,1}$. Figure 4c shows V-I curves of the Fe-JoFET in *state 0* and *state 1*. To simplify, we set $I_{READ}$ = ($I_{S,0}$ + $I_{S,1}$)/2. In *state 0*, the voltage drop across the superconducting Fe-JoFET is 0 V ($V_{STATE,0}$), while in *state 1*, the measured voltage is $R_{N,1} \times I_{READ}$ ($V_{STATE,1}$). We would like to note that $V_{STATE,1}$, and consequently the difference $V_{STATE,1}$ - $V_{STATE,0}$, can be increased by increasing the $I_{READ}$ value up to $I_{READ} = \lim_{\varepsilon \to 0}(I_{S,0} - \varepsilon)$, where $\varepsilon$ represents a safe margin from the *state 0* switching current, and is related to thermal fluctuations.

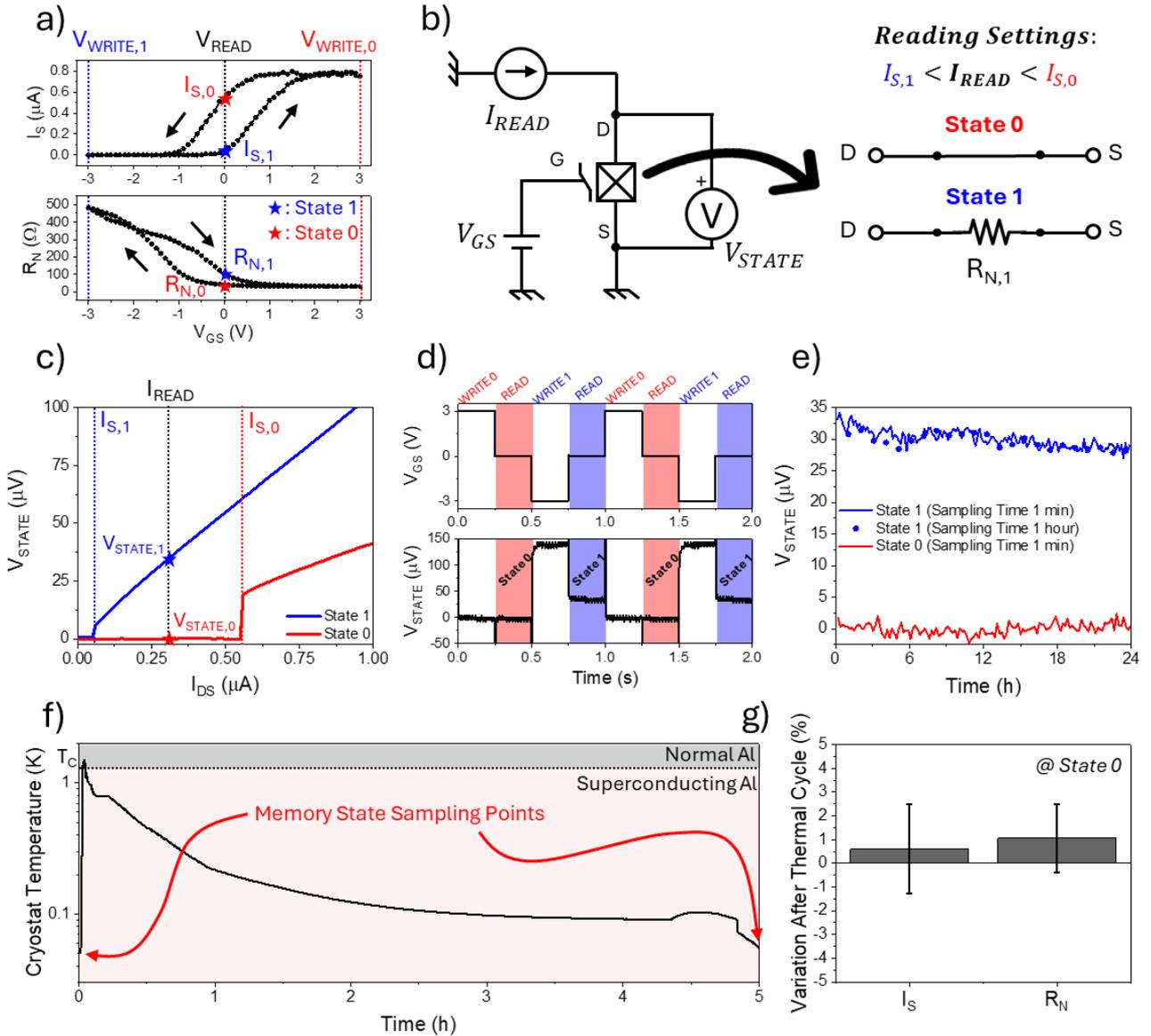

**Figure 4: Single memory cell with the InAsOI-based Fe-JoFET.** a) Switching current (top) and normal-state resistance (bottom) vs. gate voltage characteristic in *continuous $V_{GS}$* mode, highlighting writing and reading voltages and JoFET states. b) Schematic usage of the gate-programmable JoFET and equivalent electrical model in the encoded digital states. c) Voltage vs. current curve of the JoFET in the encoded digital states. d) Time-resolved usage of the JoFET along several write and read cycles. (top) $V_{GS}$ applied defining read and write phases; (bottom) $V_{STATE}$ values recorded during all the usage phases. e) Time-endurance estimation monitoring the time-resolved $V_{STATE}$ in the encoded digital states. f) Time-resolved temperature of the mixing chamber plate of the cryostat where the Fe-JoFET is located. g) Percentage variation of the switching current and the normal-state resistance after the warm-up cycle reported in (f). The measurement bath temperature is 50 mK.

Figure 4d reports the time-resolved usage of the programmable Fe-JoFET single memory cell. The $V_{GS}$ was changed between write and read values, and the $V_{STATE}$ was recorded to collect the Fe-JoFET behavior. The reproducibility of the JoFET set and read operations is observed across several writing and reading cycles, where the real-time read voltages agree with those achieved in the V-I curves reported in Figure 4c. Eventually, we evaluated the time endurance of the encoded states, a fundamental benchmark for ferroelectric behavior and a memory cell (Figure 4e). After writing the state, $V_{STATE}$ was recorded over 24 hours of measurement, varying the sampling time from 1 minute (1440 single memory accesses) to 1 hour (24 single memory accesses) to infer how much the read operation impacts the programmable Fe-JoFET time-endurance. Regardless of the sampling time, the two encoded voltage states are perfectly separated over the whole measurement. Similar results are also obtained for $I_S$ and $R_N$ over time (Figure S4). In *state 0*, the Fe-JoFET exhibits tiny variation in $V_{STATE}$, $I_S$, and $R_N$ values over 1 day. Differently, when the JoFET is programmed in *state 1*, over 24 hours, we calculated variations in $V_{STATE}$, $I_S$, and $R_N$, from 31.9 to 28.8 µV, from 103.6 to 89.7 Ω, and from 0.04 to 0.07 µA, respectively. The observed results support the non-volatility claim of the observed behavior in time, a fundamental aspect of a single memory cell. Eventually, we evaluated the single memory cell persistency against the Fe-JoFET temperature. *State 0* was encoded at 50 mK, and values of $I_S$ and $R_N$ were collected as benchmarks for memory persistence. Then, the cryostat was warmed to 1.5 K, which is higher than the critical temperature of Al, i.e., 1.2 K [25][29], and then cooled down again to 50 mK over 5 hours (Figure 4f). No significant change in the encoded values of $I_S$ and $R_N$ was observed across the thermal cycles, as reported in Figure 4g, where the percentage variation of $I_S$ and $R_N$ after the thermal cycles was calculated as $\left(1 - \frac{X_{T=0s}}{X_{T=5h}}\right) \times 100$. This is related to the fact that, regardless of the superconducting or normal state of the Fe-JoFET, the ferroelectricity acts on the number of majority carriers available in the semiconducting channel. This is a crucial remark confirming that the Fe-JoFET can be operated at the lower thermal stage of a dilution refrigerator. Yet, it is also able to retain information at higher temperatures, leading to a temperature-fault-tolerant memory cell against temperature oscillations or, in the worst case, cryostat faults.

The Fe-JoFET also represents a promising future perspective as a fully non-dissipative single memory cell by imposing the reading condition $0 < I_{READ} < I_{S,1} < I_{S,0}$. With this setting, the Fe-JoFET can be programmed in two non-dissipative states, each with a distinct Josephson inductance value ($L_J$) for *states 0* and *1*. By assuming $I_{READ} \ll I_{S,1}$, the Josephson inductance is obtained from the linearized relationship between $I_S$ and the phase drop across the JJ, namely, $L_J = \frac{\hbar}{2eI_S}$, where $\hbar$ is the reduced Planck constant, and $e$ is the elementary charge. $L_{J,0}$ = 0.6 nH and $L_{J,1}$ = 10.0 nH are calculated based on the achieved values of the switching current. Then, by flowing a non-dissipative AC current $I_{READ}(t) = I_{READ}\cos(2\pi f t)$, where $f$ is the signal frequency, a voltage $V_{STATE}(t) =$ $2\pi f L_J \times I_{READ} \cos(2\pi f t + \frac{\pi}{2})$ drops across the Fe-JoFET with a zero active power dissipation $P = \frac{1}{T} \int_0^T V_{STATE}(t) \times I_{READ}(t) \, dt = 0$.

**Conclusions**

In this work, we demonstrated the ferroelectric behavior of a Josephson Field-Effect Transistor (JoFET), leading to the development of a ferroelectric JoFET (Fe-JoFET) operating at cryogenic sub-Kelvin temperatures. The three-terminal device was fabricated on the InAs-on-insulator (InAsOI) platform, using $HfO_2$ as the gate insulator, which led to the device its ferroelectric properties. The Fe-JoFET shows evident hysteresis in both the switching current and normal-state resistance transfer characteristics, which can be precisely controlled by the applied gate voltage range. Additionally, the device can be reset to its original state on demand through a tailored ambipolar gating protocol. Engineering the Fe-JoFET with specific well-known ferroelectric insulators, as the recently discovered Si or Zr-doped $HfO_2$, is identified as the way to proceed with the reliable very large-scale integration of Fe-JoFETs. Here, we successfully operated the Fe-JoFET as a cryogenic superconducting single memory cell, where the two logic states are distinguished by their dissipative and non-dissipative behaviors. Non-volatility was confirmed over a 24-hour measurement period. Notably, the Fe-JoFET maintains its memory state even after warm-up at temperatures above the superconducting critical temperature, enabling fault-tolerant operation against thermal fluctuations. This represents a significant technological milestone in the development of robust cryogenic memory systems. The memory writing process of the Fe-JoFET closely resembles that of conventional ferroelectric memories, ensuring compatibility with existing multiplexing protocols and enabling straightforward scaling to large memory arrays. Notably, the Fe-JoFET consumes significantly less energy compared to traditional approaches, a benefit that can be further enhanced with the proposed inductive readout scheme. These findings lay a promising foundation for energy-efficient, scalable, and resilient cryogenic memory architectures.

**Acknowledgments**

This work was partially supported by H2020-EU.1.2. - EXCELLENT SCIENCE - Future and Emerging Technologies (FET) under grant 964398 (SuperGate), by HORIZON.3.1 - The European Innovation Council (EIC) – Transition Open programme under grant 101057977 (SPECTRUM), by the Italian National Research Council (CNR) under grant DFM.AD002.206 (HELICS), and by the Piano Nazionale di Ripresa e Resilienza, Ministero dell'Università e della Ricerca (PNRR MUR) Project under Grant PE0000023-NQSTI.

**Supporting Information**

**The Ferroelectric Superconducting Field Effect Transistor**

Alessandro Paghi[1*], Laura Borgongino[1], Elia Strambini[1], Giorgio De Simoni[1], Lucia Sorba[1], and Francesco Giazotto[1]

[1]Istituto Nanoscienze-CNR and Scuola Normale Superiore, Piazza San Silvestro 12, 56127 Pisa, Italy.

[*]Corresponding authors: alessandro.paghi@nano.cnr.it

**Summary**

1. Supporting Figures
2. Materials and Manufacturing Methods
   2.1. Materials and Chemicals
   2.2. InAsOI Heterostructure Growth via Molecular Beam Epitaxy
   2.3. InAsOI-Based Josephson Field Effect Transistor Fabrication
   2.4. Sample Bonding via Wire Wedge Bonding
3. DC Electrical Characterization of InAsOI-Based Josephson Field Effect Transistors

# 1. Supporting Figures

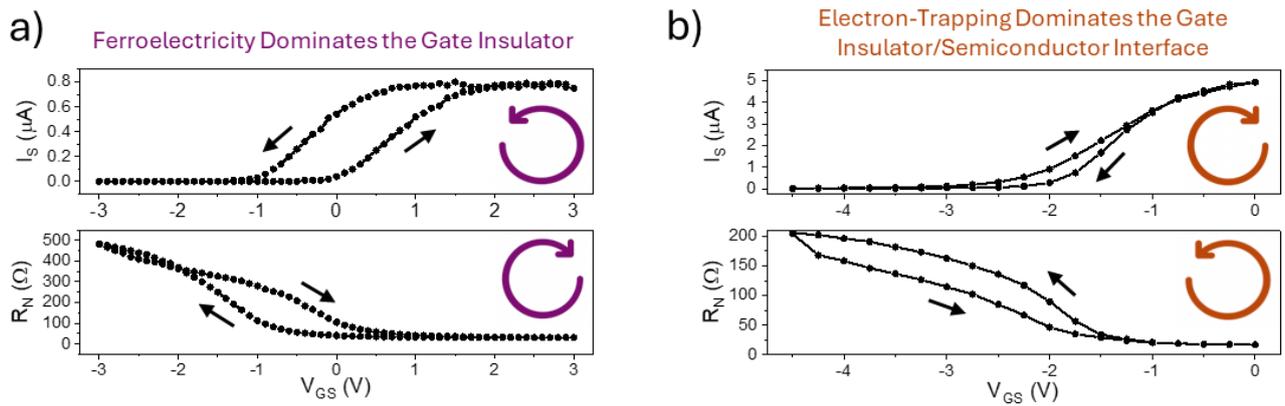

**Figure S1: a,b) Switching current (top) and normal state resistance (bottom) vs. gate voltage transfer characteristic of a JoFET where (a) ferroelectricity dominates the gate insulator or (b) electron trapping dominates the gate insulator-semiconductor interface. The measurement bath temperature is 50 mK.**

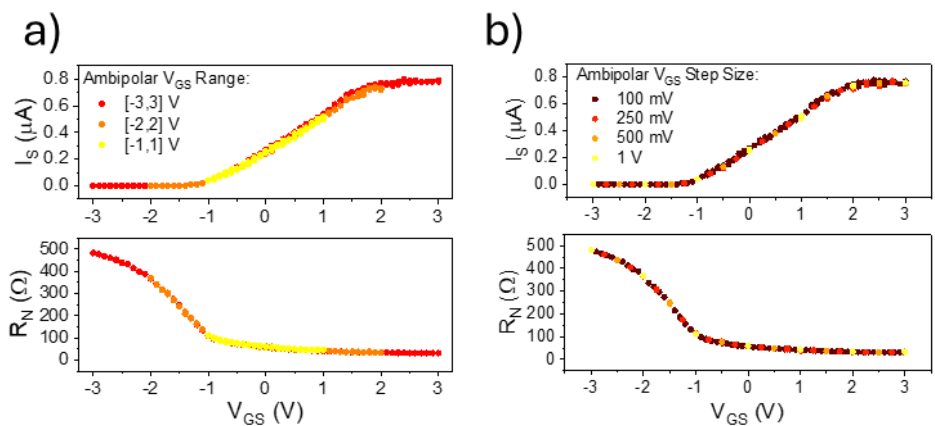

**Figure S2:** Switching current (top) and normal state resistance (bottom) vs. gate voltage transfer characteristic recorded with the *ambipolar* $V_{GS}$ routine for different scan ranges (a) and different step sizes (b).

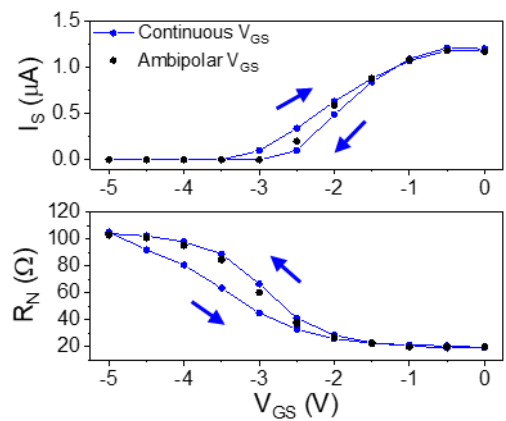

**Figure S3:** Switching current (top) and normal state resistance (bottom) vs. gate voltage transfer characteristic recorded with both *continuous* and *ambipolar* $V_{GS}$ routine for a JoFET where the hysteresis is dominated by the charge-trapping mechanism.

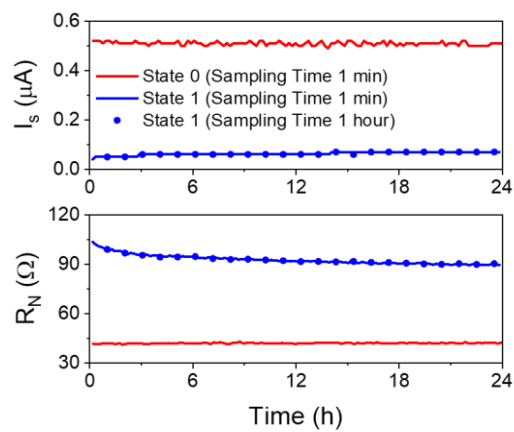

**Figure S4: Time-endurance estimation monitoring the time-resolved switching current and normal state resistance encoded digital states.**

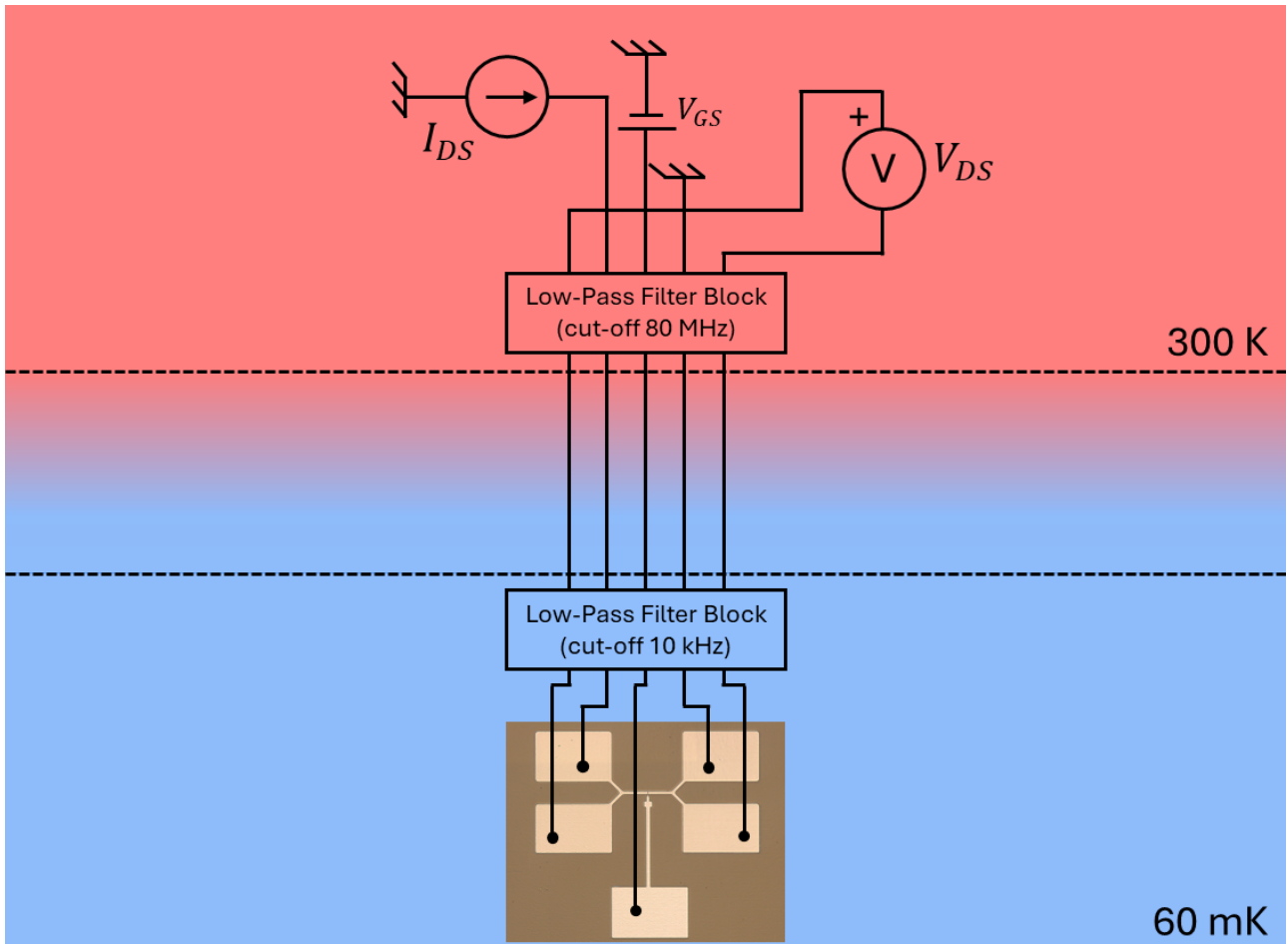

**Figure S5: DC measurement setup used to characterize the JoFETs in a dilution cryostat.**

## 2. Materials and Manufacturing Methods

### 2.1 Materials and Chemicals

GaAs wafers (2'' diameter, (100) orientation, $\rho=5.9\times10^7$ $\Omega\times$cm) were purchased from Wafer Technology LTD. Materials used for the Molecular Beam Epitaxy growth (Gallium 7N5, Aluminum 6N5, Indium 7N5, and Arsenic 7N5) were purchased from Azelis S.A.. Acetone (ACE, ULSI electric grade, MicroChemicals), 2-propanol (IPA, ULSI electric grade, MicroChemicals), S1805 G2 Positive Photoresist (S1805, Microposit, positive photoresist), AR-P 679.04 (AllResist, positive e-beam resist), MF319 Developer (MF319, Microposit), AR 600-56 Developer (AR 600-56, AllResist), AR600-71 (AllResist, remover for photo- and e-beam resist), Aluminum Etchant Type D (Transene), Phosphoric acid ($H_3PO_4$, Sigma Aldrich, semiconductor grade ≥85% in water), Hydrogen peroxide ($H_2O_2$, Carlo Erba Reagents, RSE-For electronic use-Stabilized, 30% in water), Nitrogen ($N_2$, 5.0, Nippon Gases) was provided by the Clean Room Facility of the National Enterprise for nanoScience and nanotechnology (NEST, Pisa, Italy). Diammonium sulfide (($NH_4)_2S$, Carlo Erba Reagents, 20% in water) was provided by the Chemical Lab Facility of the NEST. Sulfur pieces (S, Alfa Aesar, 99.999% pure) were purchased from Carlo Erba Reagents S.r.l. Aluminum pellets (99.999% pure) were purchased from Kurt J. Lesker Company. Aqueous solutions were prepared using deionized water (DIW, 15.0 M$\Omega\times$cm) filtered by Elix® (Merck Millipore) provided by the Clean Room Facility of the NEST.

### 2.2 InAsOI Heterostructure Growth via Molecular Beam Epitaxy

InAsOI was grown on semi-insulating GaAs (100) substrates using solid-source Molecular Beam Epitaxy (MBE, Compact 21 DZ, Riber). Starting from the GaAs substrate, the sequence of the layer structure includes a 200 nm-thick GaAs layer, a 200 nm-thick GaAs/AlGaAs superlattice, a 200 nm-thick GaAs layer, a 1.250 μm-thick step-graded $In_XAl_{1-X}As$ metamorphic buffer layer (with X increasing from 0.15 to 0.81), a 400 nm-thick $In_{0.84}Al_{0.16}As$ overshoot layer, and a 100 nm-thick InAs layer.

The GaAs layer and the GaAs/AlGaAs superlattice below the $In_XAl_{1-X}As$ buffer layer are grown to planarize the starting GaAs surface and to reduce surface roughness caused by the oxide desorption process. The metamorphic buffer layer consists of two regions with different misfit gradients df/dt. The first $In_XAl_{1-X}As$ region is composed of twelve 50 nm-thick layers with X ramping from 0.15 to 0.58, corresponding to a misfit gradient of 5.1% μm$^{-1}$. The second $In_XAl_{1-X}As$ region consists of twelve 50 nm-thick layers, with X increasing from 0.58 to 0.81, corresponding to a misfit gradient of approximately 3.1% μm$^{-1}$. The Al flux was kept constant during the growth of the buffer layer, while the In cell temperature was increased at each step without interruption. Therefore, the In concentration of the buffer is continuously changing with two different slopes. The $In_XAl_{1-X}As$ buffer layer and InAs epilayer were grown at optimized substrate temperatures of 320 °C ± 5 °C and 480 ± 5 °C, respectively. The As flux was adjusted during growth to maintain a constant group V/III beam flux ratio of 8 throughout the process. A growth rate of 0.96 μm/h was used to grow the thin InAs layer [1][2].

## 2.3 InAsOI-Based Josephson Field Effect Transistor Fabrication

In the following steps, all wetting procedures were performed on cleaned glass Beckers using stainless steel tweezers with carbon tips. Teflon-coated tweezers were used for all the steps requiring acid or base solutions. The Julabo TW2 was used to heat the solution to a specific temperature.

*Superconductor Deposition on InAsOI*

InAsOI substrates were cut into square samples (7×7 mm×mm) and sonicated in ACE and IPA for 5 min to remove GaAs dust. The air-exposed InAs surface was etched from native InAs oxide ($InAsO_X$) and passivated with S-termination by dipping the InAsOI samples in a $(NH_4)_2S_X$ solution (290 mM $(NH_4)_2S$ and 300 mM S in DIW) at 45°C for 90 s. The S-terminated InAsOI samples were then rinsed twice in DIW for 30 s and immediately loaded (~ 70 s exposure time in the air) into the load-lock vacuum chamber of an e-beam evaporator (acceleration voltage 7 kV). Samples were transferred into the deposition chamber, where a 100-nm-thick Al layer was deposited at a rate of 2 A/s at a residual chamber pressure of 1E-8 ÷ 5E-9 Torr.

*InAsOI MESA Fabrication*

After Al deposition, a layer of S1805 positive photoresist was spin-coated at 5000 RPM for 60 s (with a spin coating acceleration of 5000 RPM/s) and soft-baked at 115 °C for 60 s. The resist was then exposed via direct writing UV lithography (UVL, DMO, ML3 laser writer, λ=385 nm) with a dose of 60 mJcm$^{-2}$, resolution of 0.6 μm, high exposure quality, and laser-assisted real-time focus correction to define the MESA geometry. Unless otherwise stated, all rinsing steps were performed at room temperature (RT, 21°C). The UV-exposed samples were developed in MF319 for 45 s with soft agitation to remove exposed photoresist, then rinsed in DIW for 30 s to stop the development and dried with $N_2$. The exposed Al layer was removed by dipping the sample in Al Etchant Type D at 40 °C for 65 seconds with gentle agitation, then rinsed in DIW for 30 seconds to halt the etching, and dried with $N_2$. The exposed InAs epilayer was etched by dipping the samples in a $H_3PO_4:H_2O_2$ solution (348 mM $H_3PO_4$, 305 mM $H_2O_2$ in deionized water) for 60 s with gentle agitation, then rinsed in deionized water for 30 s to stop the etching, and dried with $N_2$. Eventually, the photoresist was removed by rinsing the InAsOI samples in ACE at 80 °C for 5 minutes, followed by IPA for 60 seconds, and then dried with $N_2$. At the end of this step, the width ($W_{JJ}$) of the JoFET was set to 5 µm.

*Markers Deposition for Aligned Steps*

After MESA fabrication, a layer of AR-P 679.04 positive e-beam resist was spin-coated at 4000 RPM for 60 seconds (with a spin coating acceleration of 10,000 RPM/s) and soft-baked at 160 °C for 60 seconds. The resist was then exposed via UVL-marker-aligned e-beam lithography (EBL, ZEISS, Ultra Plus) with a dose of 350 µCcm$^{-2}$, voltage acceleration of 30 kV, aperture of 7.5 or 120 µm, line step size of 1 nm or 200 nm, to define EBL markers for the next alignment steps. The electron-exposed samples were developed in AR 600-56 for 90 seconds with gentle agitation to remove the exposed e-beam resist, then rinsed in IPA for 30 seconds to halt the development, and dried with $N_2$. The samples were loaded into a thermal evaporator (Sistec prototype) where a 10/50-nm-thick Ti/Au bilayer was deposited at a rate of 1-2 A/s at a residual chamber pressure of 2E-5 mbar. The deposited film was lifted-off in AR600-71 at 80 °C for 5 min with strong agitation, then rinsed in IPA for 60 s, and dried with $N_2$.

*Aligned Josephson Junction Fabrication*

After EBL marker definition, a layer of AR-P 679.04 positive e-beam resist was spin-coated at 4000 RPM for 60 s (spin coating acceleration of 10000 RPM/s) and soft-baked at 160 °C for 60 s. The resist was then exposed via marker-aligned EBL (ZEISS, Ultra Plus) with a dose of 350 µCcm$^{-2}$, voltage acceleration of 30 kV, aperture of 7.5 µm, and line step size of 1 nm to define the Josephson junction length ($L_{JJ}$). The electron-exposed samples were developed in AR 600-56 for 90 seconds with gentle agitation to remove the exposed e-beam resist, then rinsed in IPA for 30 seconds to halt the development, and dried with $N_2$. Subsequently, the exposed Al layer was removed by dipping the sample in Al Etchant Type D at 40 °C for 75 seconds with gentle agitation, followed by rinsing in DIW for 30 seconds to halt the etching process and drying with $N_2$. At the end of this step, the interelectrode separation ($L_{JJ}$) and the gate length of the JoFET were set to 600 nm.

*Selectively Gate Insulator Deposition*

Samples were loaded into the vacuum chamber of an Atomic Layer Deposition system (ALD, Oxford Instruments, OpAL) where the high-*k* dielectric was selectively grown at 130 °C. We deposited a single layer of $HfO_2$, using Tetrakis(ethylmethylamino)hafnium (TEMAH) and $H_2O$ as oxide precursors. Ar was used as the carrier gas. Ar bubbling was also involved to increase the volatility of TEMAH. After reaching a base pressure of ~3-4 mTorr, the chamber pressure was increased to ~360 mTorr by injecting Ar. The deposition process follows 4 steps:

(i) TEMAH dose: TEMAH valve on; Ar bubbler: 250 sccm; Ar purge: 10 sccm; step duration: 0.9 s.

(ii) TEMAH: purge: TEMAH valve off; Ar purge: 250 sccm; step duration: 110 s.

(iii) $H_2O$ dose: $H_2O$ valve on; Ar purge: 10 sccm; step duration: 0.03 s.

(iv) $H_2O$ purge: $H_2O$ valve off; Ar purge: 250 sccm; step duration: 90 s.

We performed 250 ALD cycles to achieve a total insulator thickness of ~31 nm [3]. Eventually, the chamber was pumped to reach a base pressure of ~3-4 mTorr and then vented using $N_2$. The entire process takes ~15 h.

*Aligned Metallic Gate Deposition and Gate Stack Lift-Off*

The samples were mounted in an e-beam evaporator (acceleration voltage, 7 kV), where a self-aligned 60-nm-thick Al layer was deposited at a rate of 2 Å/s with a tilt angle of 53° at a residual chamber pressure of $1 \times 10^{-8} \div 1 \times 10^{-9}$ Torr. The deposited $HfO_2$/Al bilayer was lifted off in AR600-71 at 80 °C for 5 min with strong agitation, then rinsed in IPA for 30 s, and dried with $N_2$.

## 2.4 Sample Bonding via Wire Wedge Bonding

All the fabricated samples were provided with bonding pads ranging from $150 \times 150$ to $200 \times 200$ μm×μm and then used to connect the device with the chip carrier. Samples were glued using a small drop of AR-P 679.04, then left dry at RT for 1 hour on a 24-pin dual-in-line (DIL) chip carrier. Samples were bonded via wire wedge bonding (MP iBond5000 Wedge) using an Al/Si wire (1% purity, 25 μm wire diameter), leaving the user-bonder and the DIL chip carrier electrically connected to ground.

## 3. DC Electrical Characterization of InAsOI-Based Josephson Field Effect Transistors

Electrical characterization of JoFETs was carried out by measuring 4-wire V-I curves at 50 mK. The sample was mounted in contact with the mixing chamber (MC) plate of the Leiden CF-CS81-1400 cryostat. The electrical configuration of the measurement setup is shown in Figure S5. Source-drain current ($I_{DS}$) was injected by applying an increasing DC voltage (Voltage Source, YOKOGAWA GS200) over an input series resistor (R= 1 MΩ) at least 100 times larger than the total resistance of the remaining measurement setup. The voltage drop across the probe contacts ($V_{DS}$) was amplified (Voltage Amplifier, DL Instruments 1201, Gain = 10k, High pass filter = DC, Low pass filter = 100 Hz) and read (Multimeter, Agilent, 34410A, NPLC = 2). The gate-source voltage ($V_{GS}$) was changed between -3 and 3 V (SMU, Keithley, 2400) while the gate leakage current was collected (NPLC = 2).

The switching current ($I_S$) was estimated as the last applied current in the V-I curve before reading a voltage drop that differed from the noise floor. At the same time, the normal state resistance ($R_N$) was evaluated as the angular coefficient of the V-I linear best fitting curve for $I>I_S$.

For the $I_S$ vs. $V_{GS}$ and $R_N$ vs. $V_{GS}$ curves, two gating routines were used:

(i) *Continues $V_{GS}$*, for which a triangular waveform $V_{GS}$ was used with amplitude continuously decreasing then increasing with time (Figure 3a inset).

(ii) *Ambipolar $V_{GS}$*, for which an alternate sign $V_{GS}$ was used with magnitude first increasing then decreasing with time (Figure 3b inset).

Before recording any $I_S$ vs. $V_{GS}$ and $R_N$ vs. $V_{GS}$ curves, the JoFET was previously subjected to a polarized reset cycle using the *Ambipolar $V_{GS}$* routine in the range of -3 to 3 V.

For the time-resolved measurements, the voltage drop across the probe contacts ($V_{DS}$) was amplified (Voltage Amplifier, DL Instruments 1201, Gain = 10k, High pass filter = DC, Low pass filter = 100 Hz) and real-time collected (Oscilloscope, Tektronix, TDS2024B, averages = 128).